\documentclass[aps,prl,superscriptaddress,twocolumn,showpacs,preprintnumbers,amsmath,amssymb]{revtex4}
\usepackage{color}
\usepackage{array}
\usepackage{amsmath}
\usepackage{amssymb}
\usepackage{epsfig}
\usepackage{graphicx}
\usepackage{wasysym}
\usepackage{bbm}

\begin{document}
\title{Anomalous Aharonov-Bohm conductance oscillations from topological insulator surface states}
\author{Yi Zhang}
\author{Ashvin Vishwanath}
\affiliation{Department of Physics,
University of California at Berkeley, Berkeley, CA 94720, USA}
\affiliation{Materials Sciences Division,
Lawrence Berkeley National Laboratory, Berkeley, CA 94720, USA}

\date{\today}

\begin{abstract}
We study transport properties of a topological
insulator nanowire when a magnetic field is applied along its length. We
predict that with {\em strong} surface disorder, a characteristic signature of
the band topology is revealed in Aharonov Bohm (AB)
oscillations of the conductance. These oscillations have a component with anomalous period $\Phi_0=hc/e$, and with
conductance maxima at odd multiples of $\frac12\Phi_0$, i.e. when the AB phase for surface
electrons is $\pi$. This is intimately connected to the band topology and a surface curvature induced Berry phase, special
to topological insulator surfaces. We discuss similarities and differences from recent experiments
on Bi$_2$Se$_3$ nanoribbons, and optimal conditions for observing this effect.

\end{abstract}
\maketitle There has been much recent interest in Topological
Insulators (TIs), three dimensional solids which are insulating in the
bulk but display protected metallic surface states (see
\cite{HasanKane} for reviews),  in the presence of time reversal (${\mathcal
T}$) invariance.
Surface sensitive
experiments such as ARPES and STM \cite{HasanKane}, have confirmed
the existence of this exotic surface metal. However, so far there
have been no transport experiments verifying the topological nature
of the surface states. Besides the fact that bulk impurity bands may
contribute significantly to conductivity, the key transport property
 is the absence of localization when ${\mathcal T}$
invariance is present. This has been hard to convert into a
clear-cut experimental test. Here we discuss a topological feature of TI surface states that can
lead to a transport signature.

Consider a wire of topological
insulator, with magnetic flux applied along its length. The surface
states can be considered a collection of one dimensional modes, that
come in pairs moving up and down the wire. Time reversal  symmetry
is present at zero flux, and also, approximately for the surface
states, when the surface encloses an integer multiple of
$\frac12\Phi_0$ ($=hc/2e$) flux quanta. However, there is an important difference
between even and odd multiples of $\frac12\Phi_0$-flux. Odd multiples of $\frac12\Phi_0$-flux
leads to a $\pi$ Aharonov-Bohm phase for surface electrons, resulting in an odd
number of pairs of modes. This one dimensional state is topologically protected and cannot be localized with ${\mathcal T}$
symmetry\cite{KaneMele}. In contrast, even multiples of $\frac12\Phi_0$-flux leads to an even number of modes which are not protected. Thus, with sufficiently strong disorder,
even flux leads to a fully localized state for a long wire while odd
flux leads to a metallic state whose conductance approaches
$2e^2/h$, under ideal conditions. Interestingly, as discussed below,
a crucial component of this even-odd effect is a surface curvature induced
$\pi$ Berry's phase. Throughout we assume low temperatures so the thermal
dephasing length exceeds the sample dimensions.

Note, this oscillatory dependence has a flux  $hc/e$ period, in
contrast to Aharonov-Altshuler-Spivak (AAS) oscillations which have period
$hc/2e$. There has been much discussion  on the question of $hc/e$
vs $hc/2e$ oscillations, in mesoscopic rings\cite{Webb, Stone} and
cylinders \cite{Altshuler, Spivac}. For rings, both periods are
observed, the first period is attributed to AB
interference of single electrons, and the second to
AAS oscillation arising from weak localization
effects. However in metallic cylinders, only the $hc/2e$ period has
been experimentally reported \cite{Altshuler}. The $hc/e$ period has
been theoretically predicted, \cite{Stone, Spivac}, but always occurs with random sign, i.e.
they can peak at either even or odd multiples of $\frac12\Phi_0$.
Therefore an ensemble average tends to wash out this effect, which,
according to Ref. \cite{Stone, Webb}, is why the $hc/2e$ effect is
more commonly observed.

In contrast, the $hc/e$ oscillation
described here are unique in having maxima always at odd-integer multiples of $\frac12\Phi_0$, and only occur in strong topological insulators.

A recent experiment on topological insulator Bi$_2$Se$_3$ nanowires
has indeed reported such an anomalous $hc/e$ flux period
\cite{stanfordgroup}. However, there is a crucial difference from
the effect described above - the conductivity is found to be minimum
at the locations of the predicted maxima. Hence Ref.
\cite{stanfordgroup} is presumably observing different physics,
explaining which is an interesting open question. The regime
described in this paper is best accessed by going to strong disorder
on the surface, or by enhancing the one dimensional nature of the
system, eg. by considering narrower wires. We believe this regime
could well be accessed by future experiments of a similar nature.

We first describe the physical ingredients that give rise to this
anomalous AB effect, such as curvature induced Berry phase, in clean systems. Subsequently we report the result of numerical
experiments on disordered cylinders of topological insulators,
realized in a three dimensional lattice model. Strong disorder is confirmed to
expose this anomalous AB effect.

\begin{figure}
\includegraphics[width=0.40\textwidth] {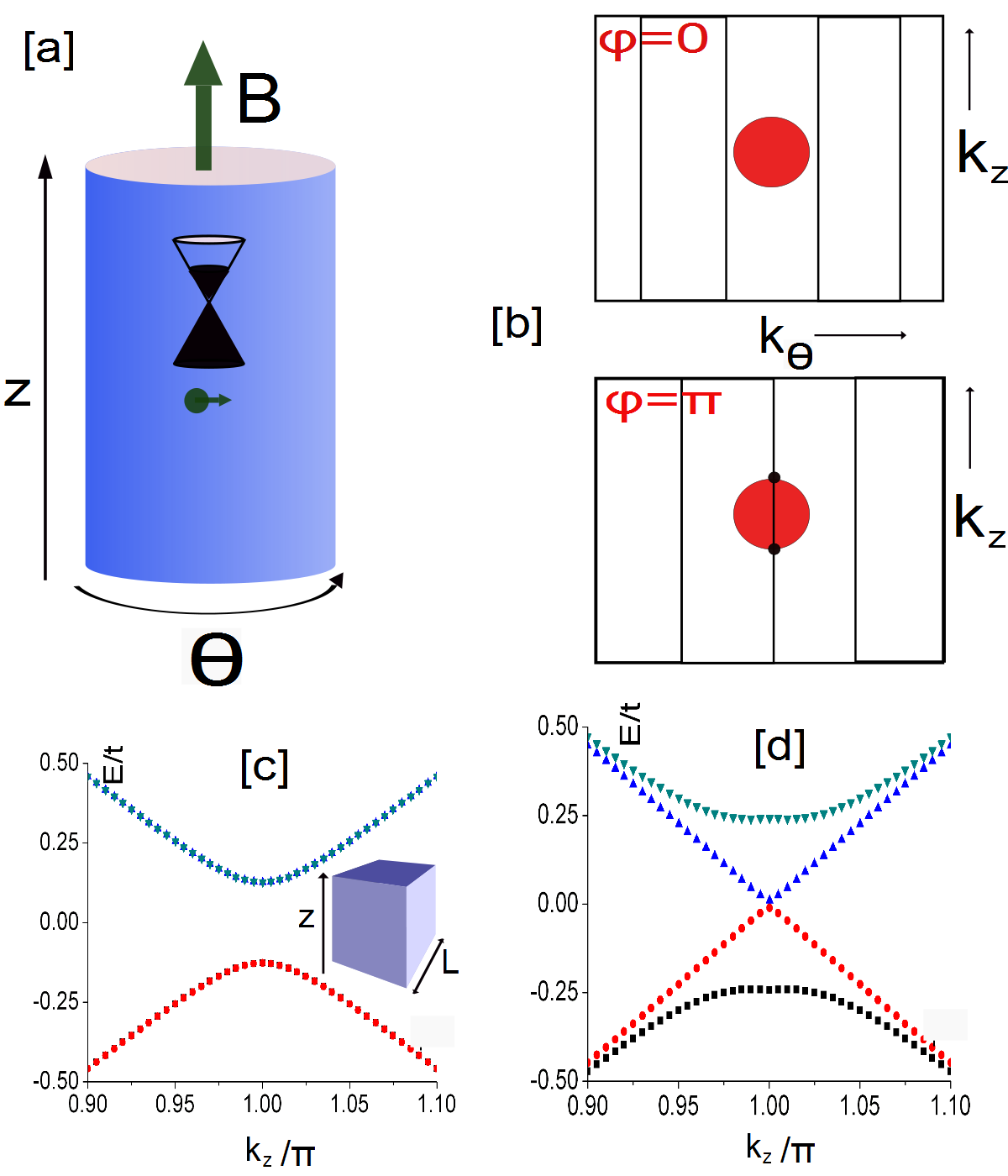}
\caption{(a) proposed geometry, with (weak) magnetic field applied along the wire axis, with flux $\varphi=2\pi\Phi/\Phi_0$ enclosed. (b) Quantization of transverse momenta $k_\theta$ for a cylinder, as a function of applied flux. The shaded area boundary in the Fermi surface. For $\varphi=0$, an even number of 1D mode pairs occurs; but with $\varphi=\pi$ flux, an odd number of pairs, protected by time reversal symmetry, is expected. Demonstration in a microscopic model:  spectrum of a cuboid of clean topological insulator, with cross section 10x10 unit cells (geometry shown in inset), as a function of momentum $k_z$ along its length.  (c) Spectrum for $\varphi=0$, each band shown is doubly degenerate (d) For flux $\varphi=\pi$, all modes except the linear one are doubly degenerate; hence an odd number of 1D mode pairs is present.}
\label{fig1}
\end{figure}

A single Dirac cone is the simplest model of TI
surface states, which naturally invites comparison with graphene, with a
pair of Dirac nodes centered at momenta $\pm K$, for each spin
projection. If scattering between the graphene's nodes are
neglected, one might conclude that the topological insulator surface
is simply '1/4th' of graphene. Here we point out a topological
effect that is special to TI surfaces,
connected to the fact that the surface Dirac fermions are
sensitive to spatial topology, in the same way as Dirac particles in a
curved two dimensional space. There is no analog of this for
graphene, even when it is rolled up into curved structures such as
nanotubes and buckyballs. The root of this difference can be traced
back to the physical interpretation of the Dirac matrices.
While in graphene, Dirac matrices act on an internal psuedospin space,
in topological insulators they act on physical spin, which are locked to the surface orientation. Hence surface
curvature introduces new effects in TIs as described in Ref
\cite{3DTIMott}. In particular, consider an electron circling the
cylindrical surface of a topological insulator. Due to the locking
of spin to the surface orientation, an additional Berry's phase of
$\pi$ is acquired during such a revolution. The surface modes of a cylinder appear
in pairs moving up and down the axis. A consequence
of the Berry phase is that there are an {\em even} number of these pairs.
If modes are labeled by angular momenta, $k_\theta$, these are quantized to half integers because of
the Berry phase. Thus there is no unpaired low energy Dirac mode at
$k_\theta=0$. In contrast, in carbon nanotubes, the $k_\theta=0$ mode
is present and responsible for the metallicity of eg. armchair
nanotubes. Now, if an additional Aharonov-Bohm flux of $\pi$ threads
the cylinder of topological insulator, canceling the curvature
Berry's phase, one reverts to the regular quantization i.e.
$k_\theta=0$ is allowed, and hence an odd number of one
dimensional mode pairs are present (see Fig. \ref{fig1}a,b).

{\em Surface Dirac Theory:}
Consider generalizing the Dirac Hamiltonian for a flat surface perpendicular to the $z$ direction,
$H = -iv_F\hbar (p_x\sigma_x+p_y\sigma_y)$
to the case when the surface is curved. We utilize the fact that the surface is embedded in three dimensional space, so ${\bf { \texttt{r}}}(x^1,\,x^2)$ is the three vector defining the surface location, as the coordinates $x^i$ ($i=1,\,2$) are varied. Then, ${\bf e}_i=\partial {\bf \texttt{r}}/\partial x^i$ are tangent vectors. Define conjugate tangent vectors ${\bf e}^i$, via ${\bf e}^i\cdot {\bf e}_j=\delta_{ij}$, the Kronecker Delta function. Naively, one might guess that the Dirac equation on this curved surface is just: $H_1 = -iv_F\hbar(\alpha^1 \partial_{1}+\alpha^2 \partial_{2})$, where $\partial_i=\partial_{x^i}$ and $\alpha^i={\bf e}^i\cdot{\bf \sigma}$ are the Pauli matrices along the tangent vectors. However, the actual form is a little more involved, and is most compactly written if we assume $x^i$ are {\em normal} (also called geodesic) coordinates. In these coordinates, (i) the ${\bf e}^i$ are orthonormal ${\bf e}^i\cdot{\bf e}^j = \delta_{ij}$; and (ii) the derivatives: $\partial {\bf e}^i/\partial x^i $, are along the surface normal ${\bf e}^3$ ($={\bf e}^1\times{\bf e}^2/|{\bf e}^1\times{\bf e}^2|$). The coefficients of proportionality are the principal curvatures  $\partial {\bf e}^i/\partial x^j=-\frac{\delta_{ij}}{R_i}{\bf e}^3$. Such a set of coordinates can always be found locally. In these coordinates it is readily established that there is an additional term:
 \begin{equation}
 H_D = -iv_F\hbar(\alpha^1 \partial_{1}+\alpha^2 \partial_{2}) +i\frac{\hbar v_F}{2}(\frac1{R_1}+\frac1{R_2})\beta
 \label{HDirac}
 \end{equation}
 where $\beta={\bf e}^3\cdot{\bf \sigma}$. A shortcut to obtaining it is requiring that the Hamiltonian $H_D$ be Hermitian and also anticommute with the local $\beta$ matrix (note, the matrices $\alpha^i,\,\beta$ are space dependent).

  A more formal derivation of the same result proceeds from the general form for a Dirac equation in curved space: $H_D = -i\hbar v_F(\alpha^1D_1+\alpha^2D_2)$, where $D_i$ is the covariant derivative along a pair of general coordinates $x^i$, defined as $D_i=\partial_{i} + \Gamma_i$, where $\Gamma_i$ is the spin connection, which here is given by the equation:
  \begin{equation}
  \label{ConnectionCommutator}
 [\Gamma_i,\alpha^j] = \partial_{i}\alpha^j + \Gamma^j_{ik}\alpha^k\\
   \end{equation}
    where $\alpha^k={\bf e}^k\cdot{\bf \sigma}$, and summation over repeated indices is assumed.
The $\Gamma^j_{ik}$ are the Christoffel symbols.   This can be solved explicitly to give the following simple expression:
   \begin{equation}
  \Gamma_i =-\frac12 \beta\partial_i\beta
   \label{spinconnection}
  \end{equation}
The Christoffel symbols are defined from the derivatives of the tangent vectors, projected into the tangent plane: $\left [ \partial_{i}{\bf e}^j \right ]_{\parallel}=-\Gamma^j_{ik}{\bf e}^k$ (indices take on values $\in \{1,\,2\}$)\cite{supplementary}. Now, we switch to normal coordinates. Clearly, the Christoffel symbols vanish since the derivatives of the tangent vector are now normal to the surface. Thus from Eqn.\ref{ConnectionCommutator} the spin connection is given by $\Gamma_{1,2} = \pm \frac{i}{2R_{1,2}}\alpha^{2,1}$, which leads back to Eqn. \ref{HDirac}. An alternate elegant formalism is developed in Ref. \cite{Lee}, for quantized Hall states.

Now, let us specialize to a cylindrical surface such that $x^1=z$
along the cylinder axis and $x^2=R\theta$, where $R$ is the radius
and $\theta$ is the angle around the cylinder. Now,  $R_1=\infty$
and $R_2=R$, in Eqn. \ref{HDirac}. The unitary transformation
$U=e^{i\sigma_z\theta/2}$, transforms that into the canonical form
$H'_D=-i\hbar v_F(\sigma_z\partial_z+\sigma_y\partial_\theta/R)$.
However, since the unitary transformation changes sign
$\theta\rightarrow \theta+2\pi$, the wavefunctions for the new
Hamiltonian satisfy {\em antiperiodic} boundary conditions on
circling the cylinder. Therefore, only angular momenta
$\hbar(m+1/2)$ are allowed, where $m$ is integer. Hence, the zero
angular momentum is absent, and there are an even number of  one
dimensional modes pairs. Now threading an additional $\pi$ flux, the periodic boundary conditions are restored, and
the parity of the mode pairs is reversed; see \cite{supplementary}
for a more general argument. Although the
cylinder has vanishing Gaussian curvature,
a nonzero spin connection leads to the Berry's phase of $\pi$. This topological property is also ultimately responsible for metallic
dislocation lines \cite{3DTIMott,Disloc}.

{\em Microscopic Model:} We now demonstrate this effect for a
lattice model of a strong topological insulator (which is more general than
the Dirac approximation). We use the model of Fu-Kane-Mele
\cite{FuKaneMele} on the diamond lattice

\begin{eqnarray}
H =  \sum_{ \langle a ij\rangle} t_{ij}c^\dagger_{i\sigma}
c_{j\sigma} +8i\lambda_{SO}\sum_{\langle\langle
ik\rangle\rangle}c^\dagger_{i\sigma}(\mathbf{\hat{d}}^{\,1}_{ik}\times
\mathbf{\hat{d}}^{\,2}_{ik})\cdot \bm{\sigma}s_{\sigma\sigma'}
c_{k\sigma'} \label{Eq:FuKaneDiamond}
\end{eqnarray}

Parameters are chosen to give a strong topological insulator \cite{supplementary} with bulk gap $\Delta=2t$.
A long cuboid with cross section $L\times L$ is taken along the weak
index direction of this model, and surface states are labeled by momenta $k_z$
along the long axis. A uniform magnetic flux $\Phi$ is introduced
uniformly through the cross section, denoted in units of the flux
quantum: $\varphi=2\pi\Phi/\Phi_0$. The surface spectrum is shown in
Fig. \ref{fig1}. All modes are doubly degenerate except the linearly dispersing mode in \ref{fig1} d.
Thus, even for small sizes $L=10$, the even-odd mode
effect and the gap closing at flux $\varphi=\pi$ is apparent. While
the breaking of time reversal symmetry at this flux implies there is
always a gap, this is seen to be very small, and the curves appear
$2\pi$ periodic, so time reversal
is approximately a good symmetry at these flux values. Thus, even in
the clean limit, metallic behavior  appears at odd multiples of $\pi$
flux, for a carefully tuned chemical potential near the
node. However, on raising the Fermi energy, the flux strength with
the larger number of modes oscillates between flux zero and $\pi$. A
robust response however is exposed by the presence of strong
disorder, which we discuss next.

\begin{figure}
\includegraphics[width=0.49\textwidth] {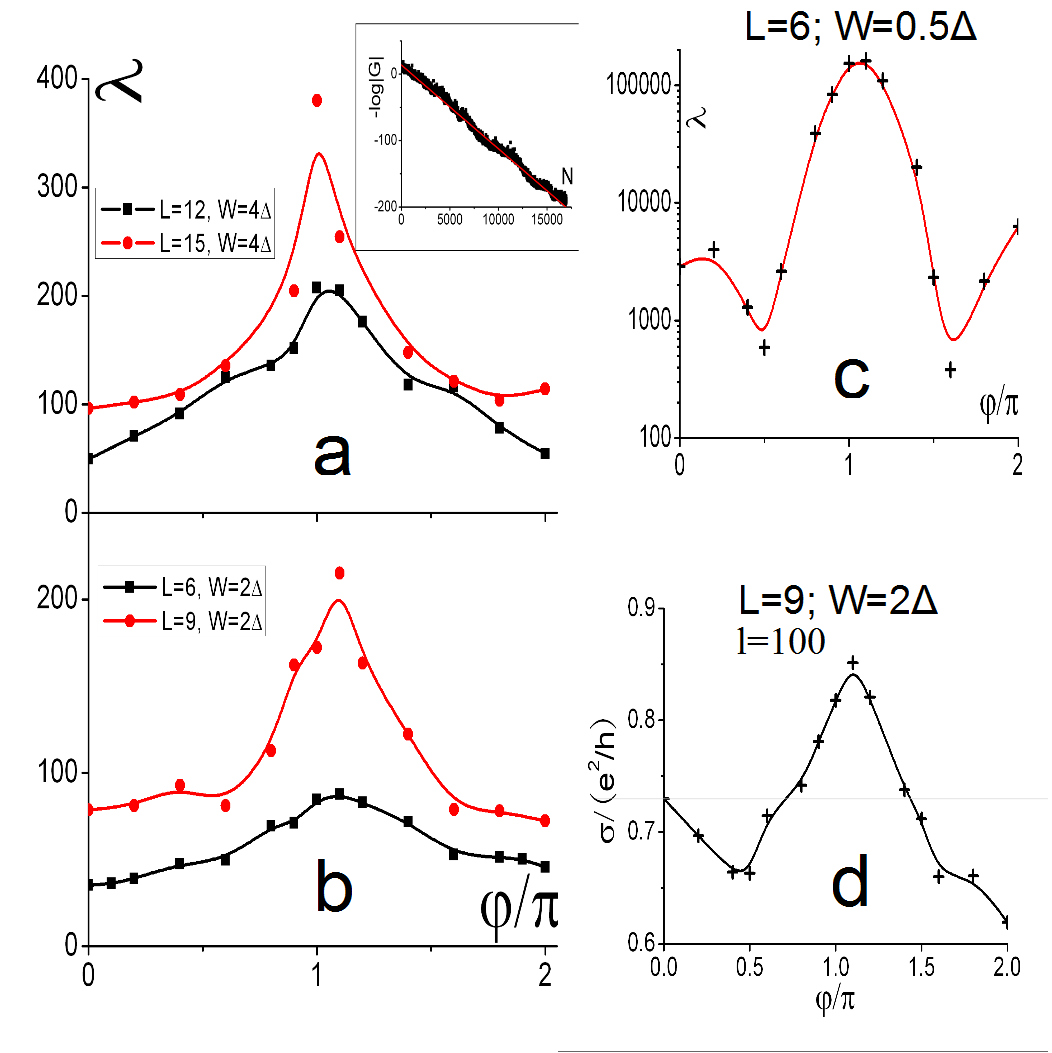}
\caption{(a), (b) and (c): Localization length $\lambda$ in the quasi one dimensional geometry, for different widths $L$ (measured in unit cells) and disorder strengths $W$ (in units of the bulk gap $\Delta$). Error bars are smaller than the symbol size. (a) and (b) Strong disorder: variation has flux-period $hc/e$ and $\lambda$ is maximum at flux $\varphi=\pi$.
 Inset: Example of the exponential decay of the Green's function used to determine the localization length.(c) Weak disorder:  Now, the $hc/2e$ oscillation period is also apparent and localization lengths are very large (note log scale here). For wires much shorter than the zero flux localization length, the $hc/e$ period will not be apparent. (d) Conductance Oscillation:  Direct measure of conductance oscillation for a wire of length $l=100$ and width $L=9$, with strong disorder $W=2\Delta$. Maximum conductance occurs at $\pi$-flux.}
\label{fig3}
\end{figure}

{\em Disordered System:} We now consider adding disorder {\em only}
on the surfaces of the model above, via a random onsite potential
$V_i n_i$, ($\mathcal T$ invariant disorder), where $n_i$ is the
charge density onsite, and $V_i$ a random variable picked from a box
distribution $[-W,\,W]$. We calculate the Greens function for a
system composed of $N$ layers in the $z$ direction, and transverse
size $L\times L$, at energy $E$: $G_N(r,r';E)$. The position and
spin coordinates are lumped into $r$. The transport properties of
the surface states are characterized by extracting the localization
length of the system in the quasi-one dimensional limit.
Subsequently, we will directly calculate the conductivity. The
technical simplification for localization length $\lambda$ is that
one only needs the Greens function between the first and last layers
of the system, as the system length is increased. Denote by the
matrix ${\bf G}^{(N+1)}_{1,N+1}$, the Greens function between the
first and last layers of a system with $N+1$ layers, where the
matrix indices refer to sites in the layer, and the energy label is
suppressed. The localization length:
$\lambda^{-1}=\underset{N\rightarrow\infty}{lim}\frac{-2}{N}log(|{\bf
G}^{(N+1)}_{1,N+1}|^{2} )$, which is self averaging, is extracted by
a linear fit (an example is shown in figure \ref{fig3}a inset) to
the logarithm of the Greens function. The latter can be efficiently
calculated \cite{Romer,supplementary}.

Results are shown in figure \ref{fig3}, where parameters were chosen to obtain a bulk gap $\Delta=2 t$, the hopping strength.
System sizes with perimeter $4L$, with $L=6,\,9,\,12,\,15$ were
studied and the quasi 1D localization length $\lambda(\varphi)$ was
extracted as a function of flux. We consider strong disorder, to
obtain a localization length short enough to be measured:
$W=2\Delta$ for the first two and $W=4\Delta$ for the last two
sizes. The chemical potential was taken to be near the middle of the
gap ($E=0$), where the Dirac node appears in the clean limit. For
these strong disorder strengths, results are  nearly {\em independent} of chemical
potential location inside the bulk gap. As seen in Figure \ref{fig3}a,b a clear maximum
in localization length is seen near $\varphi=\pi$. Note, time
reversal is already an approximate symmetry even for the
smallest sizes, given the location of the maximum near $\varphi=\pi$
and the $\varphi \rightarrow 2\pi-\varphi$ approximate symmetry.
However at $\varphi=\pi$, localization length is finite, so time
reversal symmetry breaking enters here. Clearly, for larger widths
the symmetry is more accurate, given the weaker fields prevailing on
the surface states. If we denote: $$g_\varphi=\lambda(\varphi)/4L$$
this ranges from $g_0= 1.6$ to $g_\pi= 6.3$, as the flux is varied,
for $L=15$. Hence, for these parameters, wires with aspect ratios
roughly within this range should exhibit conductance oscillations as a
function of flux with a maximum at flux $\pi$. This is explicitly checked below. We have also checked that
the {\em Zeeman} splitting induced by the field for these sizes do not
affect results qualitatively, if their energy scale $E_z \le
0.025\Delta$, which should be readily satisfied in experiments.

{\em Weak Disorder:} In Figure \ref{fig3}d, we plot a system with weak
disorder, where $W = 0.5\Delta$, and the chemical potential is tuned
to $E=0.15\Delta$, where there is a large density of states. The
localization lengths are now significantly longer (note the log
scale for $\lambda$), and a prominent anti-localization feature is
present near flux $\varphi=\pi/2,\,3\pi/2$.  The latter will
contribute to an AAS $hc/2e$ period  in conductance oscillations. Since
$g_0 = 125$, at smaller aspect ratios the system is unaware that
there is an even longer localization length at $\pi$ flux, and the
most visible feature is likely to be the $hc/2e$ AAS oscillations.
This illustrates the important role of strong disorder in observing
the effect of interest.

{\em Conductance:} Finally, in Figure \ref{fig3}d, we present the conductance of a
wire with length $l=100$, cross section $9\times9$ and strong disorder
strength $W=2\Delta$. The conductance is extracted from the Greens
function using the Kubo formula, $\sigma=\frac{e^{2}\hbar}{\pi l^{2}} Tr\left[v_{z}Im{\bf G}v_{z}Im{\bf G}\right]$ with $v_{z}=\frac{i}{\hbar}\left[H, z\right]$. The effect of leads is modeled by sandwiching the disordered system between $4\times10^3$ layers of clean wire on either side. We
take $\mu=0.15\Delta$, to obtain a finite density of states in the
leads. The localization length for these parameters is close to that in Figure \ref{fig3}b, so $l$ is intermediate between the
localization lengths obtained there. A small imaginary part
$\epsilon \sim 5 \times 10^{-4}\Delta$ is inserted in the energy to obtain finite results; results are insensitive to its precise value. Each data point is obtained by  averaging log of conductance over 500 samples. Clearly, a conductance maximum at flux $\pi$ is observed.
Note however, the $hc/2e$ oscillations are more prominent here than
in the localization length plots, and the overall contrast in
conductance is of order $0.1e^2/h$. This will increase for longer
wires and wider cross sections, where the effects of ${\mathcal T}$
breaking at $\pi$ flux are less important, and localization of
states away from this flux will set in. While a more extensive
conductance analysis is left to the future, these results
corroborate the basic picture.

{\em Scaling Analysis:}  How do these results scale to
experimentally relevant system sizes, and in particular will the
effects described here survive? As a reference, we note that a
Bi$_2$Se$_3$ nanowires studied in Ref. \cite{stanfordgroup}, was
350nm in circumference , and about 2$\mu$m in length. While a direct
conversion into unit cell lengths is difficult due to the extreme
anisotropy of the material, we estimate aspect ratios (length to
circumference) between 5-10. The circumferences themselves are about 5-10 times
larger in unit cell lengths, than those considered here. While in the truly two
dimensional limit a metallic (symplectic metal) phase is expected,
hence $g(L\rightarrow\infty)\rightarrow \infty$, this growth is
slow. The scaling to larger widths is captured by the beta function
$\beta(g)=d\log{g}/d\log{L}$, similar to the well known beta
function for conductance. This function is known for spin-orbit
metals in 2D \cite{symplecticBeta}. For large $g$, the
topological insulator surface displays identical
behavior\cite{Ryu,Brouwer}; i.e. $\beta(g)\sim 0.64/g$. We estimate
this to be reasonably accurate for $g\ge2$ for TI surfaces, then $g(\tau L)\approx g(L)+0.64\log\tau$. The localization length ratio for $\tau=10$ times wider system compared
to the $L=15$ case, will be $g(L=150)\approx 3.2$ at zero flux , an aspect ratio that
is easily exceeded. Note, the localization length at $\pi$ flux
 diverges more rapidly, since the effects of T breaking
are weaker at larger widths. It is readily seen that this scales asymptotically as $\lambda(\varphi=\pi) \sim L^4$ (since the time reversal symmetry breaking strength scales as $\delta \sim 1/L^2$, and the localization length for weak ${\mathcal T}$ breaking disorder scales as $\lambda \sim 1/\delta^2$). Thus the oscillatory effects should
remain visible at these longer scales, with strong
disorder.

{\em Experimental Realization:} In \cite{stanfordgroup}, nanowires
of Bi$_2$Se$_3$ were subjected to a field along their length and an
oscillatory dependence of conductance $O(e^2/h)$ with applied
flux was reported. The oscillation period was consistent with
surface states sensing a $hc/e$ flux. However, the conductivity
maxima were generally located at integer multiples of the flux
quantum, rather than half integer multiples as predicted here.
Therefore we believe a different effect is at play there. The main
ingredients required to access the $\pi$ flux effect discussed here
is quasi one-dimensionality and strong disorder. The former can
be achieved by studying narrower wires; eg. wires with half the
width of those studied were produced in \cite{stanfordgroup}. For
the latter, one is aiming to reduce the metallicity of the surface
layer, which in a Drude model is just $(e^2/h)k_Fl$. This can be
achieved either by decreasing the scattering length $l$, via greater
surface disorder, or tuning the chemical potential to an energy with
smaller carrier density.

{\em Conclusions:} We described a topological property of the
surface states of TIs.
An Aharonov-Bohm phase of $\pi$ strengthens the metallic nature of surface states,
leading to a clear-cut transport signature. Conductance oscillations with flux are expected in nanowires of
topological insulator, with maxima at odd-integer multiple of $\frac12 \Phi_0$. The key requirements for observing this effect are quasi
one-dimensionality and strong disorder, which we believe are
achievable given current experimental capabilities.

We thank H. Mathur, D. Carpentier, J. Moore,  and G. Paulin for insightful discussions. This work was supported by DOE grant DE-AC02-05CH11231. In independent work \cite{Jens}, similar results are obtained using an ideal Dirac dispersion with twisted boundary conditions as a simple model of TI surface states experiencing magnetic flux.

\bigskip
\center{\bf Supplementary Material}
\begin{enumerate}

\item{\it Curved Surface Dirac Theory:}
Here we remind the reader of general formalism of defining curved spacetime Dirac theories, and then specializing to our case of interest where only the space is curved. We follow reference \cite{Hansen}.

Consider flat 2+1 dimensional spacetime. We work with Euclidean time, so $x^0=\tau=it$, and $\hbar=c=1$. Then, the massless Dirac equation reads: $\left (\sum_{a=0}^2 \gamma_a\partial_a \right )\Psi=0$. The matrices satisfy the anticommutation relations: $\left \{ \gamma_a,\,\gamma_b\right \}=2\delta_{ab}$, where the right hand side is the Kronecker delta function, due to the Euclidean spacetime. The $\gamma$ matrices can just be taken as the Pauli matrices, and we can write this as a three vector ${\bf \gamma}=(\gamma_0,\,\gamma_1,\,\gamma_2)$.

Now consider a curved spacetime, defined by coordinates $x^\mu$  metric $g_{\mu\nu}$, where the indices take values $\in\{0,\,1,\,2\}$. It is always possible to define a triad of vectors ${\bf \epsilon}_\mu$, with three components each, such that $g_{\mu\nu}={\bf \epsilon}_\mu\cdot {\bf \epsilon}_\nu$. One can define also a dual frame ${\bf \epsilon}^\mu$ such that ${\bf \epsilon}^\mu\cdot {\bf \epsilon}_\nu=\delta^\mu_{\nu}$, the Kronecker delta function. (This is like the relation between bases in real space and reciprocal space, for a crystal). Now, the generalization of the Gamma matrices can be defined as: $\gamma^\mu= {\bf \epsilon}^\mu\cdot {\bf \gamma}$, or in component form $\gamma^\mu= \sum_{a=0}^2{ \epsilon}^\mu_a  \gamma_a$. The Dirac equation in curved space takes the form:
$$
\gamma^\mu(\partial_\mu+\Gamma_\mu) \Psi=0
$$
where $\partial_\mu=\partial/\partial_{x^\mu}$, and summation over repeated indices is assumed from here on. The $\Gamma_\mu$ matrices are the spin-connections, the main objects of interest here. They are given most compactly by the formula
\begin{eqnarray}
\label{spin}
\Gamma_\mu &=& \frac14\gamma_\nu \nabla_\mu\gamma^\nu\\
\nabla_\mu \gamma^\nu&=& \partial_\mu\gamma^\nu + \Gamma^\nu_{\mu\lambda} \gamma^\lambda\\
\label{covariant}
\Gamma^\nu_{\mu\lambda}&=&\frac12 g^{\nu\alpha}(\partial_\mu g_{\alpha\lambda}+\partial_\lambda g_{\mu\alpha}-\partial_\alpha g_{\mu\lambda})
\label{Christoffel}
\end{eqnarray}
where the Eqn. \ref{spin} relates the spin connection to the covariant derivative $\nabla_\mu$. This is defined via the usual Christoffel symbols, $\Gamma^\nu_{\mu\lambda}$ (Eqn. \ref{covariant}), which in turn are defined from the metric via the standard formula (Eqn. \ref{Christoffel}). The spin connection can be shown to satisfy the equation:
\begin{equation}
[\Gamma_\mu,\gamma^\nu]=\partial_\mu\gamma^\nu+\Gamma^\nu_{\mu\lambda}\gamma^\lambda
\label{commutator}
\end{equation}
which indicates that it is responsible for parallel transport of spinors.

We now specialize to the case when only space is curved. Then, we expect the metric tensor can be brought into the form $g_{00}=1;\,g_{0i}=0$, where the Latin indices $i,j,\, k\dots$ will take on values $\in \{1,\,2\}$. This can be arranged if ${\bf \epsilon}^0$ is unit modulus and orthogonal to the other two tangent vectors. Note also, the only non-vanishing Christoffel symbols are ones with all Latin indices $\Gamma^i_{jk}$. Similarly the temporal spin connection vanishes $\Gamma_0=0$. To read off the Hamiltonian, we write the Dirac equation as $\partial\Psi/\partial\tau = -H\Psi$, where $H=\gamma_0[\gamma_i(\partial_i+\Gamma_i)]$. We therefore identify  $\beta=\gamma_0$, and using $\beta^2=1$, $\gamma^i=-i\beta\alpha^i$. Substituting this into Eqn. \ref{spin}, we obtain:
  \begin{equation}
  \Gamma_i = -\frac14 (\beta\partial_i\beta +\alpha_k\partial_i\alpha^k+ \Gamma^j_{ik}\alpha_j\alpha^k )
  \end{equation}

Actually, an even simpler form for the spin connection is:
\begin{equation}
\Gamma_i = -\frac12 \beta\partial_i\beta
\label{simple}
\end{equation}
To derive this we first show (i) the spin connection is in the tangent plane, i.e. it can be expressed in terms of $\alpha_i$ and then (ii) it is given by the expression above.

To do this, its useful to use the relations\ref{commutator}. First consider the zeroth component $\gamma^0=\beta$;
$$
[\Gamma_i,\beta]=\partial_i\beta
$$
since all the relevant Christoffel symbols vanish. Now, if we can prove that $\Gamma_i$ are in the tangent plane, then the commutator can be written as $2\beta\Gamma_i=-\partial_i\beta$ which gives us the desired result Eqn. \ref{simple}. We now prove the assertion that $\Gamma_i$ lives in the tangent plane. To do this we write equations for $\alpha_i$:
$$
[\Gamma_i,\alpha^j]=\partial_i\alpha^j+\Gamma^j_{ik}\alpha^k
$$
which can be derived from Eqn\ref{commutator}, and the definition $\alpha^j=i\beta\gamma^j$. Now, if $\Gamma_i$ only has components in the tangent plane, the right hand side is proportional to $\beta$, i.e. it corresponds to the surface normal. We can check therefore that the right hand side has no components in the plane. Note, it corresponds to the vector $\partial_i{\bf e}^j+\Gamma^j_{ik}{\bf e}^k$. However, we have mentioned before that $\left [ \partial_{i}{\bf e}^j \right ]_{\parallel}=-\Gamma^j_{ik}{\bf e}^k$. Therefore the only remaining component is perpendicular to the plane, as we required.

To see this algebraically, we consider the inner product with a general in plane vector ${\bf e}_l$, which gives:
\begin{equation}
{\bf e}_l\cdot \partial_i{\bf e}^j+\Gamma^j_{il}
\label{zero}
\end{equation}
where we used ${\bf e}^k\cdot{\bf e}_l=\delta^k_l$. This can be shown to be zero. The critical input in this derivation is that
 $\partial_i{\bf e}_j=\partial_j{\bf e}_i$, which is only true because these are tangent vectors, derived by differentiating the three dimensional coordinates of the surface ${\bf e}_i=\partial{\bf \texttt r}/\partial x^i$.  

We note identical results are obtained if we choose the basis vectors as a fixed linear combination of tangent vectors: i.e. if ${\bf e}'_1=\cos \theta {\bf e}_1 + \sin\theta {\bf e}_2$ and  ${\bf e}'_2=-\sin \theta {\bf e}_1 + \cos\theta {\bf e}_2$. In particular, a natural choice is $\theta=\pi/2$, which gives a Dirac Hamiltonian in flat space of the form $H=-iv_F\hbar(\sigma_y\partial_x-\sigma_x\partial_y)$.

 We now discuss a general argument which fixes the half integer offset
of momentum quantization, and goes beyond the particular Dirac model
chosen. Note, by time reversal invariance fixes this offset to be
$\pi$ or zero. In the latter case, a gapless one dimensional mode is
present that is protected by time reversal symmetry. Now consider a
slightly different geometry, an annular cylinder with radii
$R_1,\,R_2$. Take $R_2\rightarrow\infty$ so we now have a hollow
inner cylinder. If we now shrink the inner radius $R_1\rightarrow
0$, one just obtains the bulk topological insulator. Hence, we must
have that the modes on the inner cylinder were gapped, which implies
$\pi$ phase shift. Note, although the
cylinder is 'flat' in the sense it has vanishing Gaussian curvature,
a nonzero spin connection leads to the Berry's phase of $\pi$. It is
in this sense that we refer to this phase as being surface curvature
induced - strictly it refers to a topological property of the
surface (i.e. the possibility of looping around the cylinder while
avoiding the inserted magnetic field).

\item{\it Model:} We use the Fu-Kane-Mele model of topological insulators
on a diamond lattice Eq. \ref{Eq:FuKaneDiamond} where the ${\hat{\mathbf d}}^{1,\,2}_{ik}$ are nearest neighbor unit vectors connecting
a pair of second neighbor sites $ik$. We choose nearest neighbor
hooping to be strong along direction $(111)$ with strength $t_1=2t$
and the remaining three bonds to be of equal strength $t$. The spin
orbit interaction is taken to be $\lambda_{SO}=0.25t$. The
'cylinders' used for the computations are actually parallelepipeds,
with the long axis $z$ being along $(110)$, and the cross section
axes being along $(011)$ and $(101)$. Note, the single Dirac node at
the surface of the long faces is located at $k_z=\pi$. This does not
affect any of the results since this wavevector is along the
propagation direction, and can be ignored. Note, symmetry of the
diamond lattice relates the surface Dirac nodes along the two
distinct surfaces, hence they are at the same energy. A more general
situation is when there is no particular relation between the two -
however, the topological properties should however remain unchanged,
including the anomalous Aharonov-Bohm oscillations.

\item
{\it Green's Function Calculation:} The matrix ${\bf
G}^{(N+1)}_{1,N+1}(E)$, the Greens function between the first and
last layers of a system with $N+1$ layers at energy $E$, where the
matrix indices refer to sites in the layer. This can be efficiently
calculated \cite{Romer}, assuming only nearest neighbor hopping
between layers, denoted by the matrix $\bf t$ and $\bf t^{\dagger}$.
If the single layer Hamiltonian is ${\bf H}_{N+1}$, then the Greens
function within this layer when attached to the remaining $N$ layers
is just:

\begin{eqnarray}
{\bf G}_{N+1,N+1}^{\left(N+1\right)}&=&\left[\left(E+i\eta\right)-{\bf H}_{N+1}-{\bf t}^{\dagger}{\bf G}_{N,N}^{\left(N\right)}{\bf t}\right]^{-1}\\
{\bf G}_{1,N+1}^{\left(N+1\right)}&=&{\bf G}_{1,N}^{\left(N\right)}{\bf tG}_{N+1,N+1}^{\left(N+1\right)}
\end{eqnarray}

where the energy label is suppressed. The last equation gives us the
desired Greens function.
\end{enumerate}


\begin{thebibliography}{x}
\bibitem{HasanKane} M. Z. Hasan and c. L. Kane, arXiv:1002.3895v1. X.-L. Qi and S.-C. Zhang, Physics Today 63, 33 (2010). J. E. Moore, Nature 464, 194 (2010).
\bibitem{KaneMele} C. L. Kane and E. J. Mele,  Phys. Rev. Lett. 95, 226801 (2005).
\bibitem{Webb} R. A. Webb, S. Washburn, C. P. Umbach and R. B. Laibowitz, Phys. Rev. Lett. 54, 2696 (1985). V. Chandrasekhar, M. J. Rooks, S. Wind and D. E. Prober, Phys. Rev. Lett. 55, 1610 (1985)
\bibitem{Stone} A. D. Stone and Y. Imry, Phys. Rev. Lett. 56, 189 (1986).
\bibitem{Altshuler} B. L. Al'tshuler, A. G. Aronov and B. Z. Spivak, Pis'ma Zh. Eksp. Teor. Fiz. 33, 101 (1981) [JETP Lett. 33, 94 (1981)]. D. Yu. Sharvin and Yu. V. Sharvin, Pis'ma Zh. Eksp. Teor. Fiz. 34, 285 (1981) [JETP Lett. 34, 272 (1981)]. M. Gijs, C. van Haesendonck and Y. Bruynseraede, Phys. Rev. Lett. 52, 2069 (1984).
\bibitem{Spivac} V. L. Nguyen, B. Z. Spivak and B. I. Shklovskii, Pis'ma Zh. Eksp. Teor. Fiz. 41, 35 (1985) [JETP Lett. 41, 42 (1985)]. Y. Avishai and R. Horowitz, Phys. Rev. B 35, 423–426 (1987).
\bibitem{stanfordgroup} H. Peng, et al.  Nature Materials, 9, 225-229 (2010)
\bibitem{3DTIMott} Y. Zhang, Y. Ran, A. Vishwanath Phys. Rev. B  79, 245331 (2009).
\bibitem{supplementary} See Supplementary Material.
\bibitem{Lee} D. H. Lee, Phys. Rev. Lett.  103, 196804 (2009).
\bibitem{Disloc} Y. Ran, Y. Zhang and A. Vishwanath, Nature Physics 5, 298 (2009).
\bibitem{FuKaneMele} L. Fu, C.L. Kane and E. J. Mele, Phys. Rev. Lett. 98,
10680 (2007).
\bibitem{Romer}  A. Croy, R. A. Roemer, M. Schreiber, in "Parallel Algorithms and Cluster Computing", (K. Hoffman, A. Meyer, eds.) Springer, Berlin, pp. 203-226 (2006) \texttt{cond-mat/0602300.}
\bibitem{symplecticBeta} Y. Asada, K. Slevin and T. Ohtsuki, Phys. Rev.  B 70, 035115 (2004)
\bibitem{Ryu}  K. Nomura, M. Koshino, S. Ryu, Phys. Rev. Lett. 99, 146806 (2007)
\bibitem{Brouwer}  J. H. Bardarson, J. Tworzydlo, P. W. Brouwer, C. W. J. Beenakker, Phys.Rev.Lett. 99, 106801 (2007).
\bibitem{Hansen} T. Eguchi, P. Gilkey and A. Hansen, Physics Reports 66, 213 (1980).
\bibitem{Jens} J. Bardarson and J. Moore (to appear).
\end{thebibliography}
\end{document}